\documentclass[global,twocolumn]{svjour}
\usepackage{color,afterpage,mathptm,natbib,float,epsfig,times}
\journalname{Climate Dynamics}

\begin{document}

\title{Comparing modern and Pleistocene ENSO-like influences in NW
Argentina using nonlinear time series analysis methods}
\titlerunning{Comparing modern and Pleistocene ENSO-like influences}

\author{Norbert Marwan\inst{1} \and 
        Martin H.~Trauth\inst{2} \and
	Mathias Vuille\inst{3} \and
	J\"urgen Kurths\inst{1}}

\institute{Nonlinear Dynamics Group, Institute of Physics, University of 
	Potsdam, Potsdam 14415, Germany \and
        Institute of  of Earth Sciences, University of 
	Potsdam, Potsdam 14415, Germany \and
	Climate System Research Center, Department of
    	Geosciences, University of Massachusetts, Amherst, USA}
\mail{Norbert Marwan\\
\email{marwan@agnld.uni-potsdam.de}}

\date{Received: date / Revised version: date}

\maketitle

\begin{abstract}
Higher variability in rainfall and river discharge could be of major
importance in landslide generation in the north-western Argentine Andes. Annual
layered (varved) deposits of a landslide dammed lake in the Santa
Maria Basin (26$^{\circ}$S, 66$^{\circ}$W) with an age of 30,000 
$^{14}$C years provide an
archive of precipitation variability during this time. The comparison
of these data with present-day rainfall observations tests the
hypothesis that increased rainfall variability played a major role in
landslide generation. A potential cause of such variability is the
El Ni\~no/\,Southern Oscillation (ENSO). The causal link between ENSO and
local rainfall is \textcolor{red}{quantified} by using a new method of nonlinear
data analysis, the quantitative analysis of cross recurrence plots
(CRP). This method seeks similarities in the dynamics of two
different \textcolor{red}{processes}, such as an ocean-atmosphere oscillation and local
rainfall. Our analysis
reveals significant similarities in the statistics
of both modern and palaeo-precipitation data. The
similarities in the data suggest that an ENSO-like influence on local
rainfall was present at around 30,000 $^{14}$C years ago. Increa\-sed
rainfall, which was inferred from a lake balance modeling in a previous study,
together with ENSO-like cyclicities could help to explain the
clustering of landslides at around 30,000 $^{14}$C years ago.
\end{abstract}

\section{Introduction}
Climate is a major influential factor for mass movements in high
mountain regions. Increased humidity \citep{dethier96} or
increased variability in rainfall \citep{grosjean97,keefer98}
can reduce thresholds for catastrophic landsliding. 
In order to estimate the
influence of climate in a given region, the climatic conditions
during epi\-sodes with enhanced landsliding are compared with
periods without important mass movements. The precise definition
of climate scenarios at times of high rock ava\-lanche activity
helps us to define threshold values for increased landsliding.

About 30,000 $^{14}$C years ago, multiple large
rock \linebreak avalan\-ches with volumes in excess of 10$^6$ m$^3$  
occurred in the arid to semiarid intra-andean basins of north-western
Argentina \citep{strecker99,hermanns99,trauth99}. 
A potential mechanism that could have
caused this enhanced landsliding in such an environment is climate
change. Increased humidity and/or higher inter- and 
intraannual rainfall variability
results in higher river discharge and erosion in narrow valleys and 
therefore increased destabilization of mountain fronts.

The climatic conditions in NW Argentina are not well \linebreak known for the
period at around 30,000 $^{14}$C years ago. Marine and terrestrial
records from tropical and subtropical South America indeed suggest
more humid conditions 
\citep[the Minchin period between 40,000 and 25,000~$^{14}$C years ago, e.\,g.][]{hammen94, ledru96, godfrey97, turcq97} 
and a strong El Ni\~no/\,Southern Oscillation (ENSO)
\citep[e.\,g.][]{oberhaensli90,beaufort2001}. 
Various modeling studies have shown an impact of orbital forcing on ENSO
and its weakening during the ice ages \citep{clement99,liu2000}. Thus El Ni\~no events 
may be rare around
30,000 $^{14}$C years ago \citep{clement99}. \citet{beaufort2001}, however,
inferred from Coccolithophores 
production a significant occurence of the ENSO for this period, and 
\citet{tudhope2001} based on the analysis of         
annually banded corals concluded that ENSO has been a persistent component of   
the climate system over the past 130 ka.

However, the local signal of the climatic changes in NW
Argentina is still not well defined. Laminated
     sediments from a former landslide dam\-med lake in the Santa Maria
     Basin (NW Argentina, 26$^{\circ}$S 66$^{\circ}$W) 
     contain valuable information
     about the environmental conditions for the period around \linebreak
     30,000 $^{14}$C years ago \citep{trauth99, trauth2000}.
     Hydrologic modeling of this palaeo-lake indeed indicates
     significantly wetter conditions during this time compared to the
     present 
     \citep[around 10 to 15\,\% higher precipitation, ][]{bookhagen01}. 
     Linear spectral analysis of palaeo-precipitation data derived from
     annual layered (varved) lake sediments also suggest an ENSO-like
     influence on rainfall and consequently increased 
     interannual rainfall variability in river discharge and erosion 
     \citep{trauth99, trauth2000}.
However, the results of such linear
methods are often ambiguous and not appropriate, since natural
\textcolor{red}{processes} are complex, exhibit nonstationarities and are mostly
recorded as short and noisy time series. 
In fact, data gained from sedimentation processes (as colour data)
are nonstationary by their origin and
the relationship between climatic forcing and rainfall
is not expected to be linear. Linear methods are
usually unsuitable to investigate natural complex data. In addition,
these approaches do not provide any information about a change in the
climate dynamics, e.\,g.~the sign of precipitation changes related to
ENSO-like oscillations.

The aim of our work is to test the hypothesis that an enhanced
ENSO-like \textcolor{red}{influence} on local rainfall caused a temporal landslide
cluster 30,000 $^{14}$C years ago. For this purpose, we first try to
identify ENSO-like patterns in present-day precipitation data and
infer causal links between this ocean-atmos\-phere oscillation and
local rainfall. Secondly, we search for similar influences in
palaeo-precipitation data  reconstruc\-ted from 30,000 $^{14}$C year old
lake sediments. This comparison is carried out using a new method of
nonlinear data analysis, the {\it cross recurrence plots (CRP)}, which 
can be applied to short and nonstationary complex data \citep{marwan2002pla}. 
This
procedure traces similarities and differences in several measures of
complexity in both modern and past rainfall data. Significant occurrences
of the ENSO-rainfall teleconnection together with increased rainfall 
could help to explain
enhanced landsliding 30,000 $^{14}$C years ago in NW Argentina where no
major mass movements occur today.

\section{Present-day Climatic Conditions}

Summertime climate and atmospheric circulation over NW Argentina is 
largely governed by the South American monsoon system 
\citep[e.\,g.][]{zhou98}, featuring heavy precipitation, an upper-air 
anticyclone (Bolivian High) and a low-level trough (Chaco low). 
Approximately 80\,\% of the annual precipitation amount falls 
within the summer months November\,--\,February \citep{bianchi92},
associated with southward moisture transport to the east of the Andes 
through the Andean low-level jet \citep[e.\,g.][]{nogues97}. The intra-andean 
basins and valleys, separated from this low-level moisture flux 
through the north-south running eastern Andean ridge, are arid and 
receive less than\linebreak 200~mm\,yr$^{-1}$, whereas the regions to the east 
of this orographic barrier receive more than 1500~mm\,yr$^{-1}$ \citep{bianchi92}.

Due to the seasonal change in the tropospheric temperature gradient 
between low and mid-latitudes, the subtropical westerly jet extends 
further north during the winter months, reaching its northernmost 
position around 27$^{\circ}$S. The resulting wintertime mean 
westerly flow, which prevails over the study region in the mid- 
and upper troposphere, hinders regional moisture transport over 
the eastern slopes of the Andes and leads to a typically dry 
winter climate (less than 50~mm per month). 

\begin{figure}[htbp]
\epsfig{file=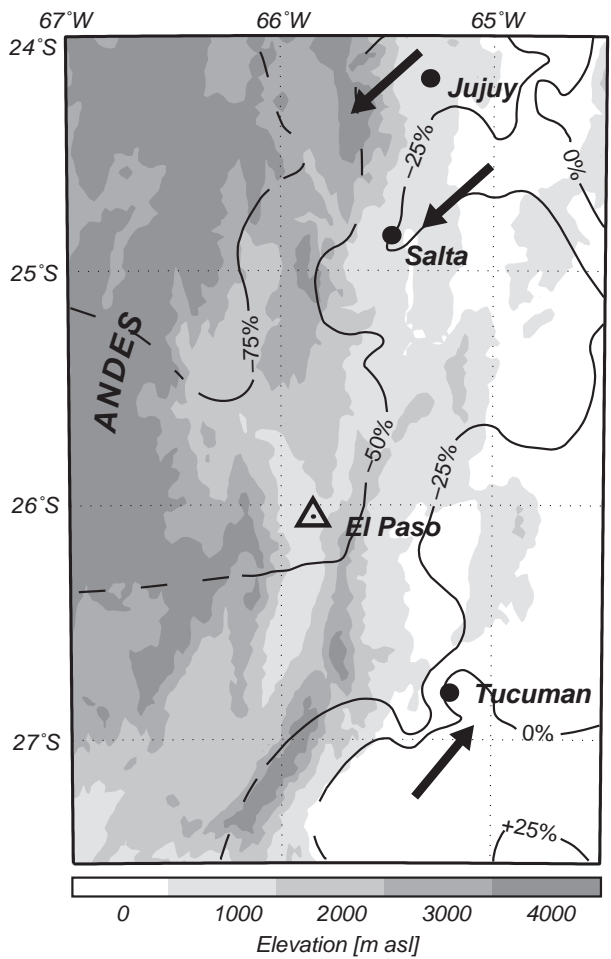, width=8cm}
\caption{Study area in the Santa Maria Basin with the locality
of annual layered lake deposits in the locality El Paso, 
the relative precipitation anomaly during the El Ni\~no 1965/66
compared to mean annual precipitation 
\citep[annual precipitation as a mean from July to June; data from][]{bianchi92} and the
prevailing wind directions during January \citep[black arrows; wind
directions from][]{prohaska76}.}\label{map_SM}
\end{figure}

\begin{figure}[htbp]
\epsfig{file=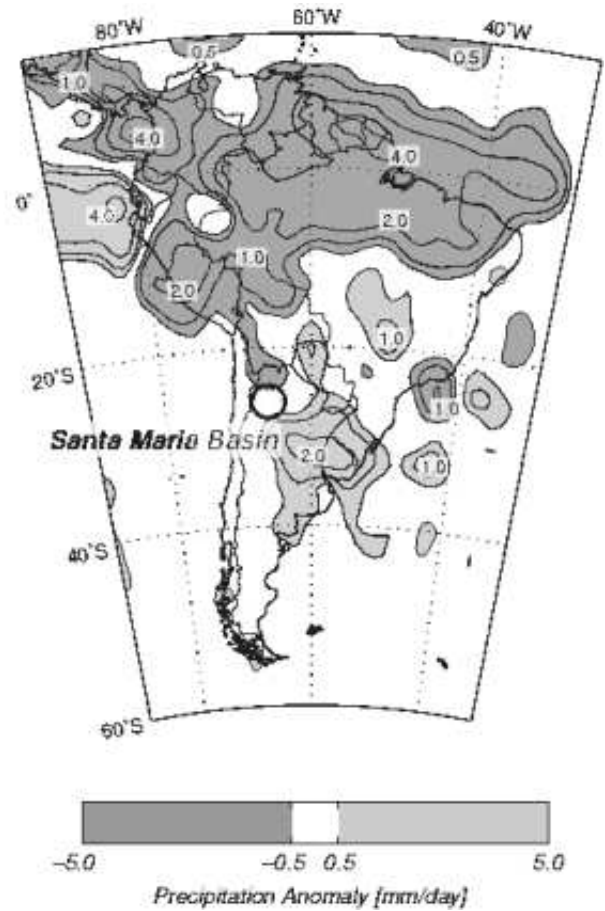, width=8cm}
\caption{Difference in precipitation (in mm day$^{-1}$) between El Ni\~no 
and La Ni\~na periods (El Ni\~no -- La Ni\~na) for December -- March 
based on CPC merged analysis of precipitation 
\citep[CMAP,][]{xie97} between 1979 and 1999. Contour interval 
is $-4$, $-2$, $-1$, $-0.5$, $0.5$, $1$, $2$, $4$ mm day$^{-1}$, regions with
difference $>0.5$ ($<-0.5$) shaded in light (dark) gray. An El Ni\~no (La Ni\~na)
event is defined as a phase of at least six consecutive months in which the
5-month running mean of SSTA in the NINO3.4 region exceeds (or remains below)
$0.5^{\circ}$C ($-0.5^{\circ}$C). The reference period for the SSTA is 1961-90.
}\label{map_SA}
\end{figure}

On interannual time scales, summer precipitation in the Central 
Andes, is primarily related to changes in meridional baroclinicity 
between tropical and subtropical latitudes, \linebreak which in turn is a 
response to sea surface temperature anomalies in the tropical 
Pacific \citep[e.\,g.][]{vuille2000,garreaud2001,garreaud2002}. 
The study region therefore shows a significant relationship with ENSO, 
featuring a weakened westerly flow with a significantly 
enhanced easterly moisture transport during La Ni\~na 
summers and strengthened westerly flow with a significantly 
subdued easterly moisture transport during El Ni\~no summers.
As a result, the rainy season is much more active during 
La Ni\~na episodes and less active during El Ni\~no episodes. 
These \linebreak[4] ENSO related atmospheric circulation 
anomalies are also evident in radiosonde data to the east of the Central 
Andes over NW Argentina (Salta), featuring enhanced southeasterly 
 \linebreak[4] (northwesterly) flow and increased (decreased) specific humidity levels 
in the lower and mid-troposphere during La Ni\~na (El Ni\~no) summers 
\citep{vuille99}. 
The notion that this ENSO influence indeed extends 
beyond just the Central Andes is further supported by several recent 
studies, which indicate that precipitation anomalies in this part of 
the Andes tend to coincide with anomalies of the same sign over SE Bolivia 
and NW Argentina \citep[e.\,g.][]{aceituno97,garreaud99}. 
\cite{bianchi92} also report a weak but significant tendency toward less rain 
during El Ni\~no years, based on a high-density network of 380 weather stations. 
\cite{trauth2000}  provide a detailed statistical analysis on the same data 
set showing that this tendency is very obvious but spatially and temporally 
highly variable. As indicated in Figure~\ref{map_SM} for the representative
El Ni\~no event 1965/66, precipitation can be decreased up to 80\,\% 
with respect to the long-term average in rainfall,
with a more significant reduction in the northern 
part of the study area. A composite analysis of the monthly summer 
precipitation (DJFM) difference between 
El Ni\~no and La Ni\~na summers between 1979 and 1999 based on CMAP satellite-derived 
precipitation data \citep{xie97} confirms this notion (Fig.~\ref{map_SA}). 
Although weak, the tendency toward less precipitation during El Ni\~no 
and more precipitation during La Ni\~na episodes is evident even 
in this low-resolution gridded data. Similar to the pattern 
in Figure~\ref{map_SM}, the ENSO signal is reversed a few degrees further south, 
where summer precipitation becomes less dominant.

\section{Methods}

It is difficult to compare rainfall proxies from 30,000 $^{14}$C year
old lake sediments with present precipitation data. The process
recording weather and climate in palaeo-archives is complex and
so far not very well understood \citep{saltzman90,bradley99}. 
Because of various signal distortions in both time and frequency 
domain, the use of linear data analysis methods reaches its limits.
The complexity of natural \textcolor{red}{processes} suggests 
the application of nonlinear methods
instead for the analysis and comparison of such complex \textcolor{red}{processes}.
Most of the nonlinear standard techniques, 
such as fractal dimensions or Lyapunov exponents, cannot be estimated
for such data \citep{kantz97}. Therefore, we have tried
to quantify cross recurrence plots of present-day and palaeo-data. 
This reveals a suite of complexity measurements which
give hints to identify similar patterns in
present-day and palaeo-data. This comparison
first tests the hypothesis that the signal
extracted from the lake sediments is an
appropriate measure for palaeo-preci\-pitation.
Secondly, it helps to reconstruct the
variability in annual rainfall as compared to
the present. Both results can then be used
to test the hypothesis that increased
interannual variability in climate can
explain enhanced landslide activity 30,000 $^{14}$C
years ago.

\subsection{Cross recurrence plots (CRP)}

An important aspect of climate changes involves nonlinear 
interactions among many components of the earth's environmental 
system  \citep[e.\,g.][]{palmer99}. These components include the oceans, 
land, lakes and continental ice sheets, and involve physical, 
biological, and chemical processes. Many of the techniques used to 
diagnose climatic variability such as Fourier analysis, 
empirical orthogonal functions or singular value decomposition are 
formulated using methods taken from linear analysis. However, 
while using these 
techniques it is difficult to analyze the nonlinear character of 
the earth's climate system. In the last two decades, a great variety
of nonlinear techniques have been developed to analyse 
data of complex \textcolor{red}{processes}. Most popular are methods to
estimate fractal dimensions or Lyapunov exponents 
\citep[e.\,g.][]{mandelbrot82,wolf85,kurths87,kantz97}. 
However, a number of pitfalls are possible due to the uncritical 
use of these methods on natural data, which are typically nonstationary 
and noisy. Furthermore, we cannot
expect low dimensions in highly complex natural \textcolor{red}{processes}.  
Therefore, we have modified and applied the
nonlinear data analysis method of cross recurrence plots,
which was recently introduced by the extension of
recurrence plots \citep{zbilut98,marwan2002npg}, for detecting
similarities and differences in the ENSO 
influence in present-day and past rainfall data. 
\textcolor{red}{In order to quantify such similarities by using
CRPs, new measures of complexity were introduced. Measures 
of complexity were developed in order to quantify the 
complexity of processes; the simplest measure is the entropy,
which distinguishs between noisy and periodically processes.
Here we use an approach which uses the geometrical structures 
which are contained in CRPs.}
This new technique is particularly efficient for the analysis of 
rather nonstationary, short and noisy data and was successfully
applied to prototypical model systems \textcolor{red}{with nonlinear interrelations}
\citep{marwan2002pla}. Thus, the CRP is an 
appropriate method for time series analysis of climate and 
palaeo-climate data.

The basic idea of this approach is to compare the dynamics of two
processes which are both recorded in a single time series. 
\textcolor{red}{Following Taken's embedding theorem \citep{takens81}, 
the dynamics of a process with $\hat m$ state parameters (i.e.~$\hat m$ 
differential equations), 
which is, however, measured by only one time series $u(t)=u_i$
with length $N$ and a sampling time $\Delta t$ (i.e.~$t=i \, \Delta t$), 
can be appropriately presented in its reconstructed phase space 
of a dimension $m$ (theoretically when $m>2\,\hat m+1$). 
Such a reconstruction can be
formed by using the time delay method (embedding), where 
for each component of the state vector a value of the time series
after a predefined delay $\tau$ (time delay) is choosen:}
\textcolor{red}{\begin{equation}\label{embedding}
\vec x_i=\left( u_i, u_{i+\tau}, \dots,  
u_{i+(m-1)\,\tau} \right),\ i = 1 \dots N,
\end{equation}}

\textcolor{red}{The dimension $m$ of such a reconstructed state or phase space 
is called embedding dimension. The time evolution of the state vectors
forms a trajectory $\vec x_i$, which runs through all possible states
at time $t=i\, \Delta t$ and, thus, present the dynamics of the process.}

The similarity in the behaviour of both \textcolor{red}{processes} in this 
reconstructed phase space can be examined by using the CRP,
which visualizes the distance between segments of their phase 
space trajectories $\vec x_i$ and $\vec y_i$ of the embedded 
time series \citep{marwan2002pla}
\begin{equation}
\mathbf{CR}^{+}_{i,\,j} = \Theta(\varepsilon-\|\vec x_{i} - \vec y_{j}\|), \quad \, 
i, j=1\dots N,
\end{equation}
where $\varepsilon$ is a predefined 
cut-off distance, $\| \cdot \|$ is the norm (e.\,g.~the Euclidean norm) and
$\Theta(x)$ is the Heaviside function. Depending on 
the type of the application, $\varepsilon$ can have a fixed
value or can vary for each $i$
in such a way that a predefined number of neighbours                     
occur within a certain radius 
$\varepsilon$ \citep{eckmann87,marwan2002npg}. 
This results in a
constant density of recurrence points in each column of the
CRP and is particularly useful in the analysis of complex
\textcolor{red}{processes} with differences in the variability of the amplitudes.

The CRP is a two-dimensional $N \times N$ array 
of points \linebreak where $N$ is the number of embedding vectors obtained 
from the delay coordinates of the input signal. The values of the 
CRP are {\it one} (black points) if trajectories lie close to each other 
(recurrence points), whereas values of {\it zero} (white \linebreak points) document 
rather large distances between two trajectories. From the occurrence 
of lines in the CRP parallel to the diagonal in the recurrence 
plot it can be seen how fast neighbouring trajectories diverge 
in the phase space. Recurrent data in a system would create 
diagonal lines in a distance $t$ from the main diagonal in such 
a plot comparing both phase-space embeddings with respect to 
the time delay $t$. It is important to note that
an additional CRP with opposite signed second time series
$\mathbf{CR}^{-}_{i,\,j} = \Theta(\varepsilon-\|\vec x_{i} + \vec y_{j}\|)$
allows to distinguish positive and negative relations.

Visual inspection of CRPs already reveals valuable 
information  about the relationship between two complex \textcolor{red}{processes}.
However, a better understanding of causal links between both 
\textcolor{red}{processes} demands a more quantitative examination of the CRPs. 
Therefore, we introduce the following two statistical measures 
of complexity \citep{marwan2002pla}: 

the {\it recurrence rate},
\begin{equation}
RR(t) = \frac{1}{N-i} \sum_{j=1}^{N-i} \left( \mathbf{CR}^{+}_{j,\,j+i} - \mathbf{CR}^{-}_{j,\,j+i} \right),
\end{equation}
and the {\it averaged diagonal length }
\begin{equation}
L(t) = \frac{\sum_{l=l_{min}}^{N-i} l\, \left[P^{+}(l,\,t) - P^{-}(l,\,t)\right]}
{\sum_{l=l_{min}}^{N-i} \left[ P^{+}(l,\,t) - P^{-}(l,\,t) \right]},
\end{equation}
where $l_{min}$ is a predefined minimal length of a diagonal line 
segment, $P^{\pm}(l,\,t)$ is the histogram
of the diagonal line lengths in $\mathbf{CR}^{\pm}$ 
at a distance $t$ from the main diagonal (i.\,e.~the time 
delay $t$ between the two phase space vectors) and $t=i \, \Delta t$. 
The $RR(t)$ is the density of adjacent states, i.\,e.~of the recurrence 
points in a CRP diagonal. $RR(t)$ therefore measures the probability 
of similar states in both \textcolor{red}{processes} after a delay time $t$. 
High densities of recurrence 
points in a diagonal cause high values of $RR$, which is typical for 
\textcolor{red}{processes} with a similar behaviour in the phase space.

Strongly fluctuating data cause short or 
absent diagonals in the CRP, whereas data from deterministic 
\textcolor{red}{processes} produce longer diagonals. If two deterministic \textcolor{red}{processes} 
have the same or similar phase-space behaviour, 
i.\,e.~the phase-space trajectory reaches the same regions 
of the phase space during certain times, then the number 
of longer diagonals will increase and the amount 
of short diagonals decrease. The average diagonal length $L$ 
measures the epoch length (i.\,e.~the time span) of significant 
similarities in the behaviour of both \textcolor{red}{processes}. The higher the
coincidence of both \textcolor{red}{processes}, the \linebreak larger the length of these diagonals. 

Consequently, high values of $RR$ and $L$ correspond to frequent 
and longer periods of similar behaviour of the \textcolor{red}{processes} as 
recorded in the time series data. Therefore, these parameters 
are appropriate quantitative measures for the 
similarities between both \textcolor{red}{processes}. However,
extrema at longer delays $t$ do not necessarily correspond
with high correlations. Future work will concentrate on the theoretical          
and more detailed investigation of the interrelations between the structures in CRPs.       

We have proposed a statistical evaluation of the 
quantitative measures of the CRP with an ensemble 
of a large amount of  surrogate
data. 

The assumption for the surrogate data is that the considered \textcolor{red}{processes}
are \textcolor{red}{linearly} independent and do not have any similar dynamics. 
These surrogates should reveal some features like in our 
original data but also 
features caused by the randomness of a possible correlation (stochastic
processes). Linear correlated noise is a paradigmatic example for 
such \textcolor{red}{processes} \citep{kantz97}. We calculate
a surrogate time series based on this class of \textcolor{red}{processes} 
with the following recursive
function, a autoregressive process of order $p$,
\begin{displaymath}
x_n=\sum_{i=1}^p a_p\, x_{n-p} + b \, \xi_n, 
\end{displaymath}

where $\xi$ is white noise and $a_i$ are coefficients which
determine the auto-correlation of the system and allow to adapt this 
stochastic system to our natural \textcolor{red}{processes}. We fit the
model to the precipitation series of the station Tucuman.
Then we perform the CRP analysis using the SOI data and the
ensemble of, e.\,g.~$10,000$ realizations of precipitation series produced 
by the AR model. Using the distributions of the $RR$ and $L$ 
measures we can estimate their empirical confidence bounds
(we will use the $2\sigma$-bounds which approximately correspond with 
the 95\% confidence level).

With these confidence bounds we can evaluate the 
obtained measures of CRP and the relations of the natural \textcolor{red}{processes}. 
Since the surrogates are
from a stationary system and the natural data are nonstationary,
we have further applied this kind of evaluation to more stationary segments
in the natural data. We got the same results. This kind of 
surrogates is a special realization, which is prototypical for linear
stochastic processes, and there are a lot of
other possibilities to construct surrogates.

\subsection{Comparison of modern and palaeo-precipitation variability}

\begin{figure}[htbp]
\epsfig{file=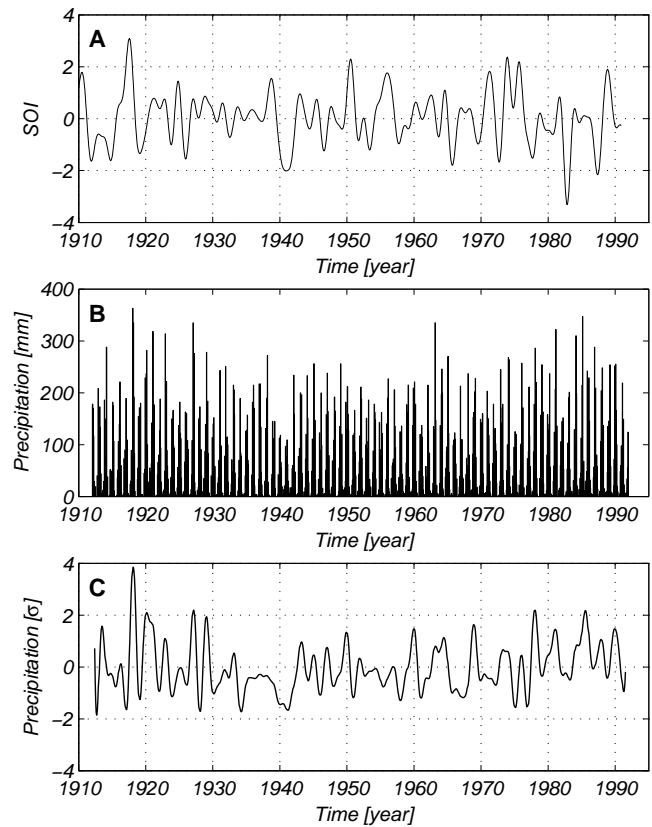, width=8.6cm}
\caption{Smoothed and $\sigma$-normalized time series of the
Southern Oscillation Index (A),
monthly precipitation data of Salta (B) and its smoothed 
and $\sigma$-normalized time series (C). SOI based on COADS data
from the NOAA 
Live Access Server (http://ferret.wrc.noaa.gov).}\label{data_SOI}
\end{figure}

In order to test the new
method on precipitation data, we first compute the CRP
for rainfall stations with well established and clear ENSO
influence. We use monthly precipitation data from the cities of Buenos
Aires (BAI) and Caracas (CAR) from the WMO data set \citep{hoffmann75}. 
For the assessment of the modern ENSO
influence on local rainfall in NW Argentina, we analyze
monthly precipitation data from three stations: San Salvador de Jujuy (JUY),
Salta (SAL) and San Miguel de Tucuman (TUC; 
Figs.~\ref{map_SM} and \ref{data_SOI}\,B). These stations in the
capitals of the provinces Jujuy, Salta and Tucuman provide the longest 
time series from this area and
are located on a north-south transect. Moreover, these locations are influenced by
different local winds; Jujuy and Salta mainly receive north-easterly and easterly
moisture-bearing winds during the summer rainy season, whereas Tucuman is
characterized by southerly and south-westerly winds \citep{prohaska76}. The Southern
Oscillation Index (SOI) is used as a measure for the ENSO variability between the years
1884 and 1990 (Fig.~\ref{data_SOI}\,A, based on COADS data). 
This index is the normalized difference between the sea level air pressure
in Tahiti and Darwin. Extreme negative values represent El Ni\~no events and extreme positive
values represent La Ni\~na events  \citep{ropelewski87}. In our analysis, 
we use monthly data, i.\,e.~twelve data points per year in order
to avoid loosing valuable intraannual information.
Moreover, longer data vectors improve the significance
of the CRP measures, thus the use
of annual data would significantly reduce the value of our results.
The potential distortion of the final result by differences in the
causal linkage between ENSO and rainfall over the year, i.\,e.~between 
dry and wet season, is believed to be of minor importance since the 
absolute values of precipitation during the dry season are low and
hence the contribution to the analysis is small. The rainfall 
variability during the dry season is not significantly different from
white noise and disappears after low-pass filtering preceding the
actual time series analysis.

\begin{figure}[htbp]
\epsfig{file=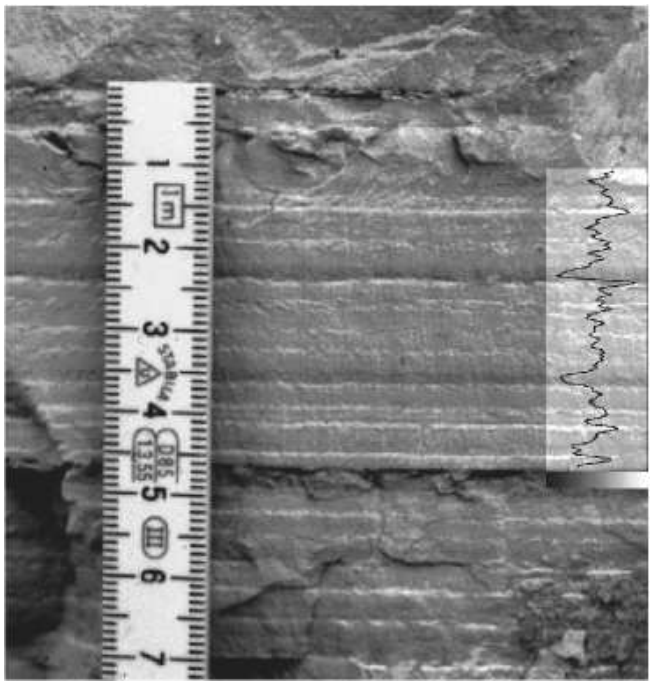, width=8.6cm}
\caption{Detail of varved lake sediments from the El Paso site in the
Santa Maria Basin with cyclic occurrence of dark red
colourations recording more precipitation and sediment
flux with ENSO-like periodicities \citep{trauth99}. The overlayed 
curve shows a representative red colour intensity transect
of the deposits.}\label{varves}
\end{figure}

The CRP analysis of the present-day ENSO and precipitation data
reveals characteristic patterns that can now be traced in 
palaeo-precipi\-tation data. 
The palaeo-precipitation variability is inferred
from varved lake sediments sampled at the location {\it El Paso} (EP; 26.0$^\circ$S, 65.8$^\circ$W)
in the Santa Maria Basin in NW Argentina (Figs.~\ref{map_SM} and \ref{varves}). 
These sediments were deposited in a landslide dammed lake 30,000 $^{14}$C years ago
\citep{trauth99,hermanns99}. Because of the internal
structure of the deposits with intra-varved changing of diatom species and the cyclic
recurrence of paired diatomite and clastic layers, these laminations are varves \citep{trauth99}.
The annual cycle with wet summers and dry winters caused
significant changes in the lake sedimentation. During the rainy season mainly
ocher coloured silty sediments were deposited; during the subsequent dry season a
thin white layer consisting of the skeletons of silica algae (diatoms) was deposited. Due
to its white colour, the diatomaceous layers can be used to identify single years in these
sediments. Recurring intense red colouration of the silty part of the annual layers is
sourced from reworked older sediments which are eroded and deposited only during extreme
rainfall events. Therefore, the intensity of red colour in the varved deposits can be
interpreted as a proxy for precipitation variation at El Paso site \citep{trauth99,trauth2000}.
The more intense the red colour the higher was the precipitation during the rainy season.

\begin{figure}[htbp]
\epsfig{file=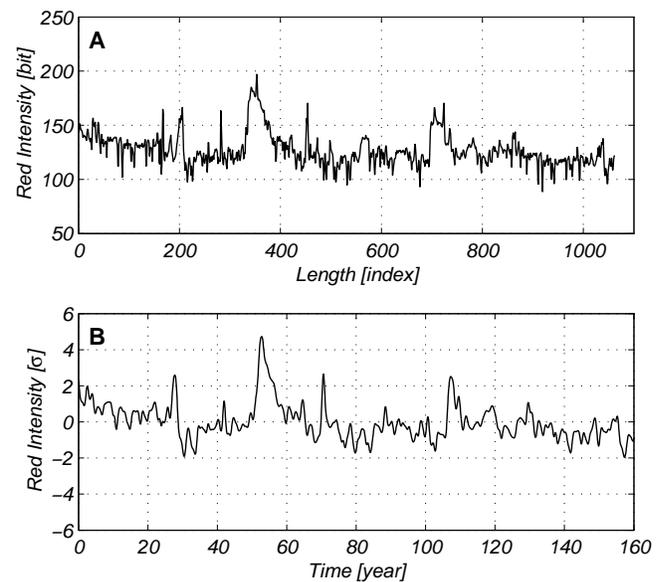, width=8.6cm}
\caption{Red intensity values of the lake sediments of site EP160
on (A) a length scale and on (B) 
a time scale and after smoothing and normalization; 
the unit of raw data is one bit, the unit of transformed
and smoothed data is the standard deviation $\sigma$.}\label{varvesdata}
\end{figure}

The colour intensity of a section of the sediments profile with a length of 160 varves
was gained by scanning high quality photographs. After digital pre-processing, a
time series of red intensity values on a length scale was obtained. We transform
these data to a time scale assuming an annual recurrence of the diatomaceous 
layers. Within single varves 12 subannual data points are computed by logarithmic 
interpolation of the data taking into account the exponential decrease of the 
sedimentation rate during the annual cycle (Fig.~\ref{varvesdata}). The power 
spectrum estimate of the red colour intensity reveals
significant peaks within the ENSO frequency band of 2 to 4 years,
suggesting an ENSO-like influence \citep{trauth2000}. Because of the 
nonstationarity of these data
(the sedimentation process in a lake is not 
stationary, resulting in nonstationary proxy parameters
for the in-lake processes; mean of the first half of the time series is $0.30$, of the second
half is $-0.32$; standard deviation of the first half of the 
time series is $1.13$, of the second half is $0.71$), linear correlation analysis is
unsuitable. Therefore, we apply the CRP analysis to these data.

\section{Results of Nonlinear Data Analysis}

First, all time series are normalized and low-pass filtered 
using a 7th-order Butterworth filter with a cutoff frequency of 1/18 month$^{-1}$
in order to remove the predominant annual cycle from the data 
(Figs.~\ref{data_SOI}\,C and \ref{varvesdata}). \textcolor{red}{Butterworth 
filters
are from the infinite-duration impulse response type (IIR filters) and
have a monotonically decreasing response with respect to frequency \citep{elliott87}.}

Next, 
the filtered rainfall data and the Southern Oscillation Index 
(SOI) are embedded into a phase space using $m=3$ and $\tau=9$. 
The method of nonlinear time series analysis using delay time 
embedding relies on a choice of good delay time and the 
embedding dimension. Proper values for these parameters 
are determined using the methods of false nearest neighbours 
and mutual information \citep{kantz97}. The quantitative analysis 
of cross recurrence plots is then applied to pairs of 
time series, local precipitation records and the Southern 
Oscillation Index (SOI). The CRPs are computed using a 
fixed amount of nearest neighbours with $\varepsilon=15$\%. 
Since the statistics of CRPs are sensitive to changes in 
the cutoff distance, we have run sensitivity tests in 
order to find the optimum value of $\varepsilon$. 
The value of 15\,\% appears to be the best choice 
receiving robust and precise results.

\begin{figure}[t]
\epsfig{file=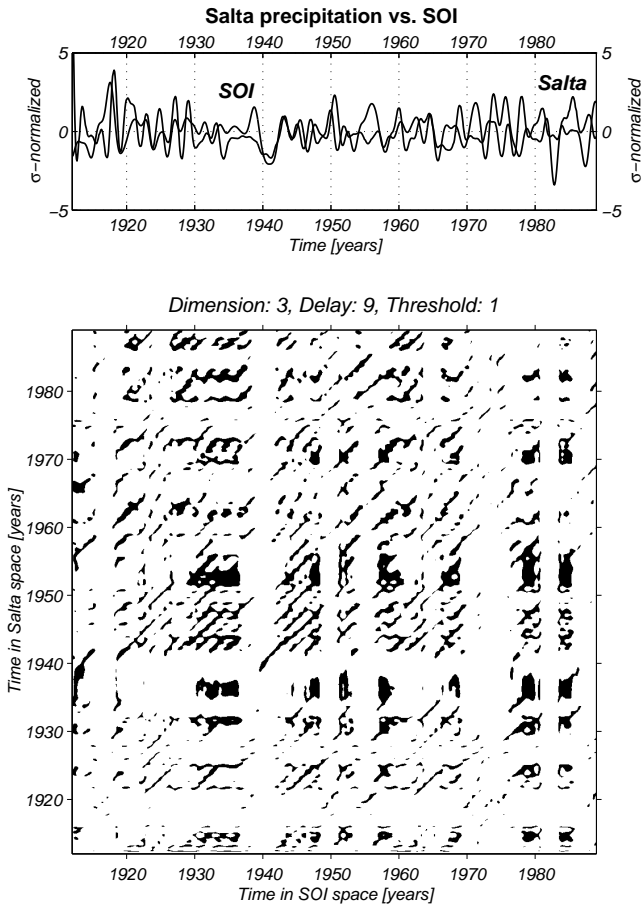,  width=8.6cm}
\caption{Cross recurrence plot of SOI vs.~precipitation data from the city
of Salta (SAL). 
The $x$-axis shows the time along the phase space trajectory of the SOI
and the $y$-axis that of SAL.
Black points represent the occurrence of similar states in both
\textcolor{red}{processes}. Diagonal lines correspond with epochs of similar dynamics in
both \textcolor{red}{processes}. The amount and length of these lines can be used as 
measures of the similarity of both \textcolor{red}{processes}.
}\label{CRPsal}
\end{figure}

\begin{figure}[t]
\epsfig{file=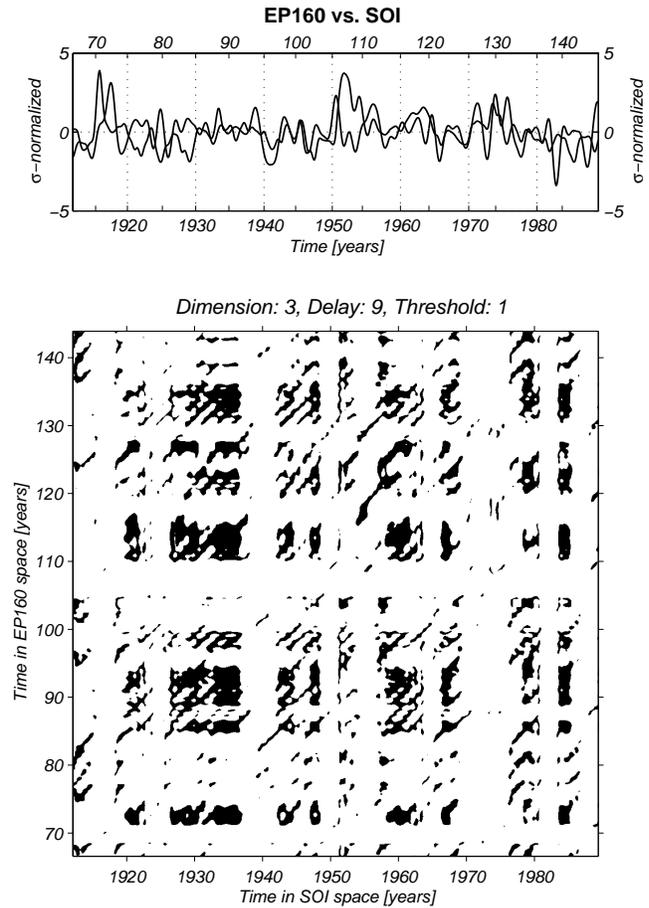, width=8.6cm}
\caption{Cross recurrence plot of 
SOI vs.~the best matching section of palaeo-precipitation (EP160). 
Scaling as in Fig.~\ref{CRPsal}.
The $x$-axis shows the time along the phase space trajectory of the SOI
and the $y$-axis that of EP160.}\label{CRPvarves}
\end{figure}

The CRPs of all pairs of time series show similar features. 
The significant similarities between CRPs obtained from \linebreak modern 
(Fig.~\ref{CRPsal}) and palaeo-precipitation data 
(Fig.~\ref{CRPvarves}) indicate that the red colour 
intensity records from the varved lake sediment do
reflect rainfall in NW Argentina. First
we discuss the CRP of Salta precipitation (data series SAL) 
vs.~the Southern Oscillation Index (SOI) and the CRP of red 
colour intensity of varves (data series EP160) vs.~SOI.
The $x$-axis represents time along the phase space trajectory 
of the SOI, whereas the $y$-axis represents the time along 
the phase space trajectory of SAL or EP160, respectively.
The CRP of SAL vs.~SOI exhibits longer
diagonal lines in two to four year intervals, which matches
the same frequency band obtained by the power spectral analysis
(Fig.~\ref{CRPsal}). 
This indicates that some sequences
of the phase space trajectory of the SOI recur in
sequences of the phase space trajectory of SAL after
relocating by the time of two to four years.
Vertical white bands in the CRP represent less frequent states in SOI,
such as horizontal white bands suggest for SAL. 
The latter occurs with intervals of more than ten years. 
The CRP between EP160 and SOI shows similar
characteristics as the CRP described above (Fig.~\ref{CRPvarves}). 
Longer diagonal lines have spacings of about two to four years. White bands 
occur at time scales of more than ten years. Some linkages
in both CRPs are obvious by visual inspection. Next, the quantitative analysis 
of the CRPs is performed in order to study statistically these relations
and to allocate the predefined causality patterns to 
certain localities.

\begin{figure}[tbp]
\epsfig{file=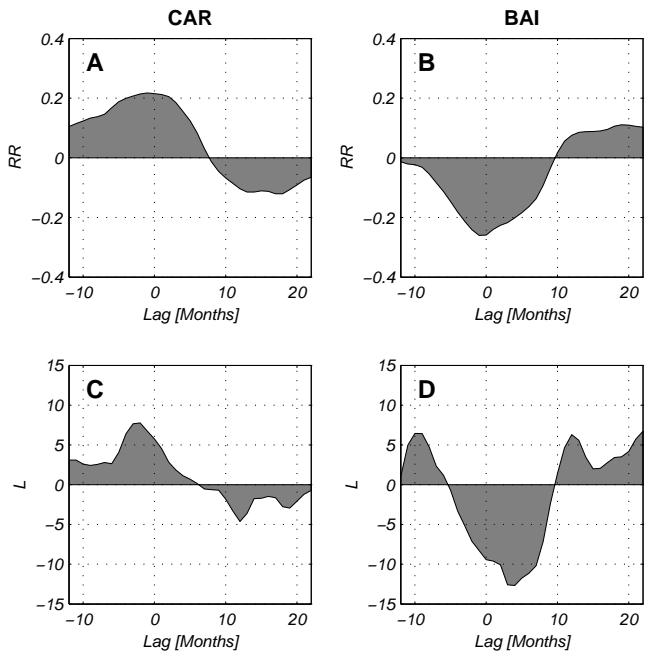, width=8.6cm}
\caption{$RR$ and $L$ measures of the cross recurrence plots  between
SOI and precipitation in Caracas (A, C) and Buenos Aires
(B, D) with a well-established and clear ENSO influence. Extreme values reveal high
similarity between the dynamics of the rainfall and the ENSO. }\label{varves_ref_crqa}
\end{figure}

In order to calculate the measures of complexity
$RR$ and $L$ between the rainfall data
and SOI, we used a delay time in the range 
between $-12$ and $+22$ months, 
i.\,e.~these measures are determined in a small corridor above and
below the main diagonal. We are interested in the extrema and in
the time lag where they occur and we get the following results for 
the various pairs of records. From an ensemble of $10,000$ realizations
of a 5th-order AR-model we calculate the $2\sigma$-bounds of 
their distributions for $RR$ and $L$. The coefficients for the
AR-model are adapted to the Tucuman precipitation (we also
used AR-models adapted to the rainfall data of the other stations,
which revealed similar results). The order of the AR-model
is determined with the Akaike's Information Criterion and a
criterion, which assesses whether the residues follow white noise 
\citep{schlittgen99}.

\begin{figure*}[btph]
\epsfig{file=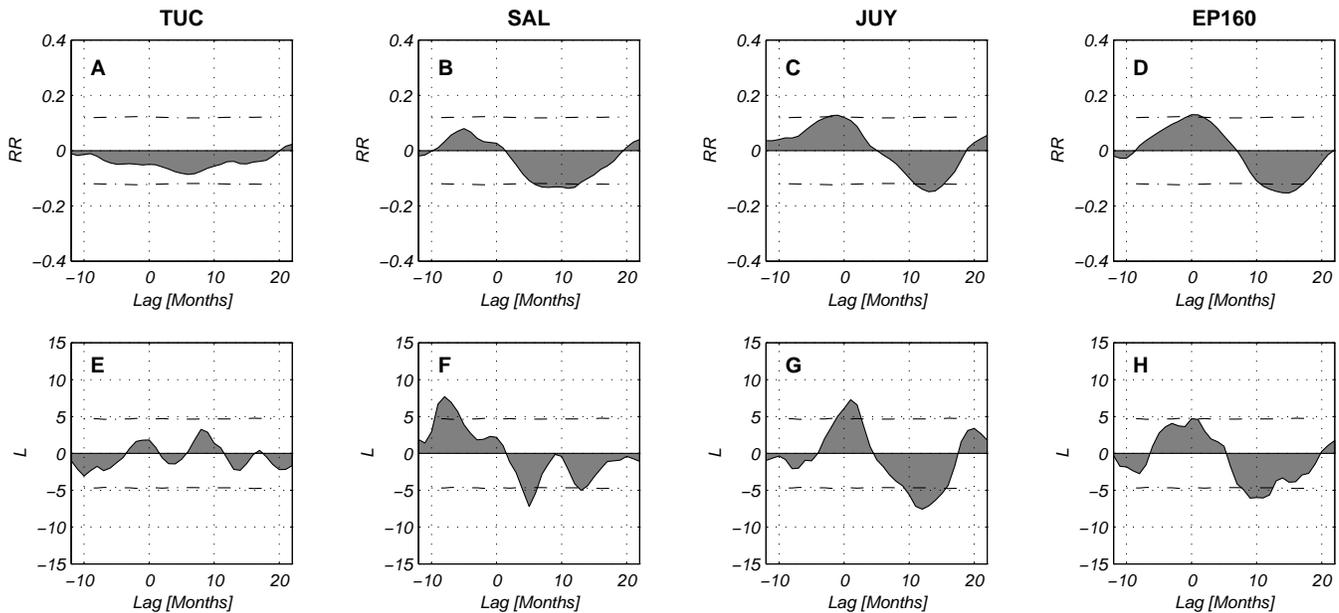, width=17.8cm}
\caption{$RR$ and $L$ measures of the cross recurrence plots between
SOI and precipitation in Tucuman (A, E), Salta (B, F),
Jujuy (C, G) and palaeo-precipitation (D, H). Extreme values reveal high
similarity between the dynamics of the rainfall and the ENSO. 
The dash-dotted lines are the empirical $2\sigma$-bounds 
from the distributions of an ensemble of data based on a
5th-order AR-model.}\label{varves_crqa}
\end{figure*}

The CRP measures between CAR and SOI reveal extreme positive
values and between BAI and SOI extreme negative values, which reflect
the strong influence of ENSO in these areas  (Fig.~\ref{varves_ref_crqa}).
The parameter $RR$ of the CRPs between TUC and SOI has small 
negative values, which do not exceed the $2\sigma$-bounds,
and does not show preferences for 
a distinct lag. The parameter $L$ has also small values, but it 
has rather small maxima and minima at delays of $-1$, $4$ and $8$ 
months. These results indicate that the precipitation in
Tucuman is not \linebreak strongly influenced by ENSO. If there
is a weak influence, the rainfall would increase during 
El Ni\~no (Fig.~\ref{varves_crqa}\,A, E). 
The analysis of JUY and SOI reveals clear positive values 
around a lag of zero and negative values after 
about 8 -- 12  months, which suggests
a significant link between Jujuy rainfall and
ENSO (Fig.~\ref{varves_crqa}\,C, G). 
The measures for the analysis SAL vs. SOI show 
smaller maxima for a delay of about zero and 
minima after a lag of about 8 -- 12 months.
We therefore infer a weaker linkage between Salta
rainfall and ENSO (Fig.~\ref{varves_crqa}\,B, F; 
the disrupted minima in the
$L$ parameter at around ten months is due to 
the short data length and a resulting
nonstationarity in the CRP).
The measures for SAL and JUY exceed the $2\sigma$-bounds.

The 30,000 $^{14}$C year old precipitation data are 
not simply comparable with present-day data, because there is no 
information available about how to synchronize the rainfall 
records with modern climate indices. 
Therefore we seek the time window in
these data showing the highest coincidence in the dynamics using 
maximum values for $RR$ and $L$ as the key criterion. 
Although the observed coincidence is not very high, it 
yields the time section in the palaeo-precipitation 
record EP160 which can be best correlated with modern data.
In our palaeo-data EP160 
we find such a section represented by maximum and minima values for
$RR$ and $L$ for delays of about zero and ten months, similar
to those found for JUY and SAL (Fig.~\ref{varves_crqa}\,D, H). 
The $RR$ and $L$ measures exceed also the $2\sigma$-bounds.

To use the minima at lags around 8 -- 12 months for climatological               
interpretations is difficult and might lead to erroneous conclusions, but       
these characteristic patterns of positive and negative
interrelations can be used to compare the present-day and
palaeo-data. The positive and negative interrelations have 
the same time delay between 10   
and 12 months in the present-day and the palaeo-data.

\section{Discussion}

We applied the method of cross recurrence plots (CRPs) to
modern and palaeo-precipitation data in order to compare the
magnitude and causes of rainfall variability in the NW Argentine 
Andes today and during the time of enhanced landsliding
at around 30,000 $^{14}$C years ago. \textcolor{red}{CRPs are able to
look for nonlinear interrelations between two processes.}
The major result
from this analysis is the significant similarity between 
the complex dynamics of modern rainfall and the palaeo-precipitation
as recorded in the red colour intensity record from the lake 
sediments in the location El Paso.
The distances between
longer diagonal lines in the CRP of both records are about
two to four years, the approximate time of recurrence of
extreme ENSO phases today. The first implication of this result 
is that the red colour intensity of the sediments is
indeed a good proxy for the rainfall intensity 30,000 $^{14}$C
ago. This \textcolor{red}{result} is in line with the observations of \cite{trauth2000}
suggesting an enhanced erosion of red-coloured clastic
sediments during heavy rainfall events today whereas precipitation 
usually only reaches the elevated areas with mainly
greenish low-grade metamorphic rocks exposed. This \textcolor{red}{effect} causes
predominant greenish to buff-coloured clays deposited in the
former lake basin \citep{trauth99,trauth2000,trauth2002}.

Since our analysis of modern data reveals a strong relation
between local rainfall in the northern part of the study
area (Jujuy) and ENSO, we interpret this similarity as an indication
of a strong ENSO-like influence in the Santa Maria Basin at
around 30,000 years. In contrast, there is no significant
linkage between the
modern rainfall in Tucuman and ENSO. This \textcolor{red}{result}
could indicate
that ENSO does not influence precipitation in the southern
part of the study area or this influence is rather diffuse
or changing in time.

The CRP between the SOI and the rainfall in Jujuy and Salta 
reveals a positive relation without any large delay, i.\,e. the 
occurrence of an El Ni\~no at the end of a year would cause
decreased rainfall in the rainy season from November to January
and the occurrence of a La Ni\~na would cause increased
rainfall during this time of the year. The opposite response
after a delay of 8 -- 12 months is not easy to interpret, 
because we do not know which mechanism actually caused
this linkage. The time span between the identified maxima and minima is 
about one year and could be explained by the fact that La Ni\~na
events often follow El Ni\~no events. 
The smooth shape of the CRP measure curves are artefacts caused by       
low-pass filtering of the time series. The measures of CRP of Tucuman precipitation and SOI
show non-significant values without any characteristic delays.
The analysis of varve data reveals a significant positive relation 
between SOI and palaeo-precipitation at the location El
Paso. Similar to the modern situation, the CRP shows a significant 
negative relation with SOI after a delay of about ten months.
Both interrelations are rather similar to those of ENSO--JUY and
ENSO--SAL.

The similarities between the \textcolor{red}{time series of} the modern
rainfall data and the palaeo-precipi\-tation record from the lake
sediments suggests that an ENSO-like oscillation was active at
around 30,000 $^{14}$C years ago (roughly corresponding to 34,000
cal.~years BP), which corresponds with the results of the
investigation of Coccolithophores production \citep{beaufort2001}.
A younger landslide cluster in the same region
at around 5000 $^{14}$C (corresponding to 5800 cal.~years BP) was
also explained by a stronger ENSO influence at that time
(\citealp{trauth2000}; palaeo-ENSO evidence from \citealp{keefer98,sandweiss2001,haug2001}). 
The spacing
between both landslide clusters is around 28,000 years. Although
two landslide clusters do not allow to infer a systematic
recurrence of such events, we believe that there is some evidence 
that these events correspond to the periods of a strong
ENSO-like \textcolor{red}{variation} as reported from deep-sea sediments off-shore Peru 
\citep{oberhaensli90}, in the Indo-Pacific Ocean \citep{beaufort2001}
and New Guinea corals \citep{tudhope2001}. These long-term
ENSO records suggest a mixed precession-glacial forcing on
ENSO resulting in significant 23- and 30-kyr cyclicities, 
which confirms
model results and recently inferred relations
between ENSO variability and insolation 
\citep{clement99,liu2000,rittenour2000}.

In the semiarid basins of the NW Argentine Andes, the ENSO-like 
\textcolor{red}{variation could have caused} significant fluctuations in local
rainfall at around 30,000~$^{14}$C years similar to the modern conditions. 
Together
with generally higher moisture levels as indicated by lake balance 
modeling results, this mechanism could help to explain 
enhanced landsliding at around 30,000 and 5,000 $^{14}$C
years ago in the semiarid basins of the Central Andes.

\section{Conclusions}

The quantitative analysis of cross recurrence plots has revealed 
similarities in the \textcolor{red}{evolution of the phase space trajectory} of 
climate indices and
present-day and past rainfall. \textcolor{red}{In comparison to the usually 
less variable climate during ice ages, our result
suggests an enhanced impact of ENSO-like 
conditions on local climate in the Santa Maria Basin
30,000 $^{14}$C years ago associated with a strong inter- and 
intraannual variability of rainfall and an intensification of 
moisture transport.} A more variable climate due to an enhanced ENSO-like 
\textcolor{red}{impact}
could have raised the risk of landsliding in this region and could
help to explain enhanced landslide activity at around 30,000 
$^{14}$C years ago.

\section{Acknowledgments}
This work is part of the Collaborative Research Center 267 
{\it Deformation Processes in the Andes} and
the Priority Programme {\it
Geomagnetic variations: Spatio-temporal structures, processes 
and impacts on the system Earth} supported by the
German Research Foundation. We gratefully acknowledge \linebreak U.~Schwarz and
M.~Thiel for useful conversations and discussions and 
U.~Bahr and M.~Strecker for support of this 
work. Further we would like to thank the NOAA-CIRES
Climate Diagnostics Center for providing COADS and CMAP data.


\bibliographystyle{springer}
\bibliography{../mybibs,varvesbib}

\end{document}